\documentclass[conference]{IEEEtran}
\ifCLASSINFOpdf
\else
\fi

\usepackage{color}
\usepackage{mylay}

\usepackage{amsmath}

\usepackage[linesnumbered,ruled]{algorithm2e}

\usepackage{xcolor}
\usepackage{lipsum}
\begin{document}
%
% paper title
% Titles are generally capitalized except for words such as a, an, and, as,
% at, but, by, for, in, nor, of, on, or, the, to and up, which are usually
% not capitalized unless they are the first or last word of the title.
% Linebreaks \\ can be used within to get better formatting as desired.
% Do not put math or special symbols in the title.
\title{The adaptive zero-error capacity for a class of channels with noisy feedback}

% author names and affiliations
% use a multiple column layout for up to three different
% affiliations
\author{\IEEEauthorblockN{Meysam Asadi, Natasha Devroye}
\IEEEauthorblockA{School of Electrical and Computer Engineering\\
University of Illinois at Chicago, Chicago IL 60607, USA\\
Email: masadi, devroye @ uic.edu}
}

% conference papers do not typically use \thanks and this command
% is locked out in conference mode. If really needed, such as for
% the acknowledgment of grants, issue a \IEEEoverridecommandlockouts
% after \documentclass

% for over three affiliations, or if they all won't fit within the width
% of the page, use this alternative format:
% 
%\author{\IEEEauthorblockN{Michael Shell\IEEEauthorrefmark{1},
%Homer Simpson\IEEEauthorrefmark{2},
%James Kirk\IEEEauthorrefmark{3}, 
%Montgomery Scott\IEEEauthorrefmark{3} and
%Eldon Tyrell\IEEEauthorrefmark{4}}
%\IEEEauthorblockA{\IEEEauthorrefmark{1}School of Electrical and Computer Engineering\\
%Georgia Institute of Technology,
%Atlanta, Georgia 30332--0250\\ Email: see http://www.michaelshell.org/contact.html}
%\IEEEauthorblockA{\IEEEauthorrefmark{2}Twentieth Century Fox, Springfield, USA\\
%Email: homer@thesimpsons.com}
%\IEEEauthorblockA{\IEEEauthorrefmark{3}Starfleet Academy, San Francisco, California 96678-2391\\
%Telephone: (800) 555--1212, Fax: (888) 555--1212}
%\IEEEauthorblockA{\IEEEauthorrefmark{4}Tyrell Inc., 123 Replicant Street, Los Angeles, California 90210--4321}}

% use for special paper notices
%\IEEEspecialpapernotice{(Invited Paper)}

% make the title area
\maketitle

% As a general rule, do not put math, special symbols or citations
% in the abstract
\begin{abstract}
 The adaptive zero-error capacity of discrete memoryless channels (DMC) with noiseless feedback has been shown to be positive whenever there exists at least one channel output ``disprover'', i.e. a channel output that cannot be reached from at least one of the inputs. Furthermore, whenever there exists a disprover, the adaptive zero-error capacity attains the Shannon (small-error) capacity.  
 Here, we study the zero-error capacity of a DMC when the channel feedback is noisy rather than perfect. We show that the adaptive zero-error capacity with noisy feedback is lower bounded by the forward channel's zero-undetected error capacity, and show that under certain conditions this is tight.
\end{abstract}

% no keywords

% For peer review papers, you can put extra information on the cover
% page as needed:
% \ifCLASSOPTIONpeerreview
% \begin{center} \bfseries EDICS Category: 3-BBND \end{center}
% \fi
%
% For peerreview papers, this IEEEtran command inserts a page break and
% creates the second title. It will be ignored for other modes.
\IEEEpeerreviewmaketitle

\section{Introduction}
% no \IEEEPARstart
Shannon determined that the zero-error capacity, denoted by $C_0$, of
a point-to-point channel whose channel $W(y|x)$ has confusability
graph $G_{X|Y}$ is positive if and only if there exist two inputs
that are ``non-confusable'' \cite{Shannon56}. Equivalently, it is
non-zero if and only if the independence number of $G_{X|Y}$ is
strictly greater than 1. 
%A multi-letter expression for the zero-error capacity $(C_0)$ of the channel with confusability graph $G_{X|Y}$ is known, and is given by the normalized limit as the block-length $n \rightarrow \infty$ of the maximum independent set of the $n$-fold strong product of $G_{X|Y}$ \cite{Shannon56}.

%Although zero-error communication scheme guarantee that no error happens, there exist numerous DMC's with independent number of $G_{X|Y}\leq 1$, and thereby the zero-error capacity $C_0$ is equal to $0$. In general, 
Shannon's condition for positive zero-error capacity $C_0$ is restrictive; that for positive zero-error capacity in the presence of perfect output feedback is less so.  
In the set of slides \cite{Massey07}, Massey showed that it is possible to communicate at a non-zero rate with zero-error over a DMC with noiseless feedback  if, and only if,  there exists at least one channel output that is reachable from some but not all the channel inputs. Such a channel output is called a ``disprover''. Not only does the existence of a disprover allow for positive rates, but Massey showed that with perfect feedback, the adaptive zero-error capacity of channels attains the small-error Shannon capacity $C$. Note that the adaptive zero-error capacity allows for adaptive and variable-length codewords rather than blockcodes. Shannon only considered block codes for zero-error feedback channels in \cite{Shannon56}. %showed that for zero-error block codes feedback does not increase the zero-error capacity. %, just like in the small error scenario. %This is not so for adaptive zero-error codes. %the zero-error feedback capacity is

The  binary erasure channel (BEC) and the Z-channel are examples of channels whose zero-error capacity $C_0$ without feedback is equal zero, but, as both contain a disprover,  have zero-error capacity equal to their Shannon capacity (positive in general) in the presence of perfect feedback.  
%In \cite{Massey07}, it was shown that positive zero-error rate for these channels are achievable using feedback. In fact, Massey proves that in general for DMC with noiseless feedback, existence of disprover is necessary and sufficient to achieve zero-error rates equal small-error Shannon capacity $C$. Thus, the zero-error capacity can be improved using feedback, while it is known that feedback does not increase small-error Shannon capacity. 
In order to achieve such zero-error rates, an adaptive communication scheme is used in which the transmitter repeatedly sends a message until it sees that it has been correctly received.
%realizes --  through the perfect feedback channel --  that it has been correctly received. %it receives a confirmation from receiver through feedback channel.     

While the zero-error capacity in the presence of feedback has not been extensively studied beyond the slides of Massey \cite{Massey07}, as we will show, it has strong connections with the zero-undetected-error capacity \cite{Forney68} with noiseless feedback \cite{Bunte12}. Two types of communication errors occur: i) erasure errors, when the decoder is unable to uniquely decode any message, and ii) undetected-errors, when the decoder uniquely decodes an erroneous message. The zero-undetectable error capacity $C_{0u}$, first considered by Forney \cite{Forney68}, denotes the maximal number of inputs that can be transmitted to ensure that the probability of an undetectable error is exactly zero. Forney derived a lower bound for the zero-undetected-capacity $(C_{0u})$  of a channel, which he showed is positive if, and only if, this channel contains a disprover. Later on, a tighter lower bound on $C_{0u}$ was derived by Ahlswede \cite{Ahlswede96}, which was shown to be tight for two classes of channels in \cite{Pinsker70} and \cite{Csiszar95}. Finally, in \cite{Bunte12} it was shown that the zero-undetected-error capacity for a channel with noiseless feedback, denoted by $C_{0uf}$, is equal to the small-error Shannon capacity $C$ if the channel contains at least one disprover. Note that in general $C_0 \leq C_{0u}\leq C_{0uf} \leq C$.  

{\bf Contribution.} 
In this paper we focus on zero-error communication for a general DMC with feedback. In Theorem 1, we detail the proof of a result outlined by Massey in his slides \cite{Massey07} for the zero error capacity of a channel with noiseless feedback, $C_{0fa}$. In Theorem 2, our main result, we consider noisy (rather than noiseless) feedback, and using an adaptive zero-error scheme, we prove that the adaptive zero-error capacity of the channel with noisy feedback, $C_{0fa}^{noisy}$, is at least the zero-undetected-error capacity of the forward channel $C_{0u}^{(f)}$.  Theorem 2  further outlines a class of channels for which this lower bound is tight. 

 \section{Definitions}
 \label{Sec2:preliminary}
 
%\subsection{Notation}
Let $x_{i}^{j}: = (x_{i}, x_{i+1}, \dots, x_{j})$ when $i \leq j$ and $|x_{i}^{j}|=j-i+1$ denote its size. For simplicity we write $x^n=x_{1}^{n}$. Let $\mathcal{B}=\{0,1\}$ be the binary set, and $\mathcal{M}$ be the message set. %Let $\gamma_n = o(n)$, and $\gamma_n \rightarrow \infty$ as $n \rightarrow \infty$ (e.g. $\gamma_n=\log (n)$).
%\subsection{Definitions}

{\bf Channels.} A channel $({\cal X}, {\cal Y}, W)$ is used to denote a generic DMC  with finite input alphabet
$\mathcal{X}$, finite output alphabet
$\mathcal{Y}$, and transition probability $W(y|x)$. We write
$W^{n}$ to denote the
channel corresponding to $n$ uses of $W$:
\[
W^{n}(y^{n}|x^{n}) = \prod\limits_{j=1}^{n} W(y_j|x_j), \quad x^{n} \in \mathcal{X}^{n}, y^{n} \in \mathcal{Y}^{n}.
\]
We consider channels with feedback, with a forward channel $(\mathcal{X}_{(f)}, \mathcal{Y}_{(f)}, W_{(f)})$ (subscript ${(f)}$) and a backward channel $(\mathcal{X}_{(b)}, \mathcal{Y}_{(b)}, W_{(b)})$ (for feedback, subscript ${(b)}$).
%We use subscript $_{(f)}$ to indicate the input/output alphabet and the channel law $W$ of the forward DMC $(\mathcal{X}_{(f)}, \mathcal{Y}_{(f)}, W_{(f)})$, and  use subscript $_{(b)}$ to indicate the  backward channel $(\mathcal{X}_{(b)}, \mathcal{Y}_{(b)}, W_{(b)})$. Wherever the corresponding DMC is obvious, we omit these subscripts.

{\bf Small error capacity $C$ without feedback.}
A $\mathcal{C}(\mathcal{M},n)$ code for DMC $W$ with message set $\mathcal{M}$ without feedback, consists of a message set ${\cal M}$ of size $2^{nR}$, for $R$ the rate and $n$ the blocklength, and encoding and decoding functions $\mathcal{F}$ and $\mathcal{G}$ respectively:
%Let $c^{(n)}(m) \in \mathcal{C}(\mathcal{M},n)$ denote a codeword corresponding to message the codeword corresponding to message $m \in \mathcal{M}$. Let $\mathcal{F}$ and $\mathcal{G}$ denote the encoding and decoding functions, respectively.
\[
\mathcal{F}: \mathcal{M} \rightarrow \mathcal{X}^n, \;\;\;\;\;\; \mathcal{G}: \mathcal{Y}^n \rightarrow \mathcal{M}.
\]
Let $c^{(n)}(m)$ denote a codeword corresponding to message $m$, i.e. $c^{(n)}(m) = F(m)$ and let
\[
\lambda_m^{(n)} = Pr(\mathcal{G}(y^n) \neq m | X^n = c^{(n)}(m) ),
\] 
be the conditional probability of error given that message $m$ was sent. The maximum and average, respectively, probabilities of error for  a $\mathcal{C}(\mathcal{M},n)$ are defined as
\[
\lambda^{(n)}= \max\limits_{m \in \mathcal{M}} \lambda_m^{(n)}, \; \;\;\;\; P_{e}^{(n)}=\frac{1}{|\mathcal{M}|} \sum\limits_{m\in \mathcal{M}} \lambda_m^{(n)}.
\]
%and the average probability of error $P_{e}^{(n)}$ is 
%\[
%
%\]
%The rate of $\mathcal{C}(\mathcal{M},n)$ code is 
%\[
%R=\frac{\log_2 |\mathcal{M}|}{n}.
%\]   
The small error capacity $C$ for channel $W$ is defined as the largest number $R$ such that there exists a sequence of $\mathcal{C}(\mathcal{M},n)$ codes such that $\lambda^{(n)}$ tends to $0$ as $n \rightarrow \infty$. 

{\bf Zero-undetected error code and capacity $C_{0u}$ \cite{Bunte12}.} A zero-undetected-error code of block-length $n$, denoted by $\mathcal{C}_{0u}(\mathcal{M},n)$, again consists of a message set ${\cal M}$, an encoding function 
\[
\mathcal{F}_{0u}: \mathcal{M} \rightarrow \mathcal{X}^n,
\]
that encodes messages $m$ to $c_{ou}^{(n)}(m)$, and a decoding function $G_{0u}$ described as follows.  
%After receiving $y^n$, the decoder lists all probable messages that might generate the corresponding output $y_1^n$. 
Let $M(y_{1}^{n})$ denote the set of probable messages corresponding to a received output $y_{1}^{n}$
\begin{equation}
M(y^{n}) = \{ m \in \mathcal{M}: W^{n}(y^{n}|c_{ou}^{(n)}(m)) > 0\}. \label{eq:probable}
\end{equation}
The decoder declares an erasure, denoted by $\mathcal{E}$, if there exist more than one possible message that could have yielded output $y^n$, i.e. $|M(y^{n})|>1$. A zero-undetected-error decoder function is then defined as
\[
\mathcal{G}_{0u}(y^n)=
\begin{cases} 
M(y^{n}) &\mbox{if } |M(y^{n})|=1  \\ 
\mathcal{E} & \mbox{if } |M(y^{n})|>1. 
\end{cases}
\]
%Let $\lambda_m^{u}$ denote the probability of an undetected error
%\[
%\lambda_m^{u} = Pr(\mathcal{G}_{0u}(y^n) = m', m' \neq m | X^n = c_{0u}^{(n)}(m) ).
%\] 
A valid zero-undetected-error code must have no undetected errors, hence the maximal error probability is given only by the probability of erasures as
%$\lambda_m^{u} = 0$. Thus,
\[
\lambda_m = Pr(\mathcal{G}_{0u}(y^n) = \mathcal{E} | X^n = c_{0u}^{(n)}(m) ).
\]
The zero-undetected capacity $C_{0u}$ for channel $W$ is defined as the largest rate $R$ such that there exist a sequence of $\mathcal{C}_{0u}(\mathcal{M},n)$ codes that $\max_{m\in {\cal M}} \lambda_m$ 
%\lambda^{(n)}$
tends to $0$ as $n \rightarrow \infty$.
%{\color{red} One can also define the zero-undeteced error capacity with feedback }

\begin{figure}
	\centering
	{\scalefont{0.6}
		\begin{tikzpicture}[xscale=0.7,yscale=0.4 ]
		\def \r {6};
		\def \s {14};

		\draw [-> , very thick] (\r-1.4,\s+0.5)--(\r-0.4,\s+.5);
		\node [align=center] at (\r-1.2,\s+1) {$m \in \mathcal{M}$};
		
		\draw [very thick] (\r-0.4,\s-0.5) rectangle (\r+2.2,\s+1.5);
		\node [align=center] at (\r+.9,\s+0.5) {Feedback Assisted\\ Encoder};
		
		\draw [-> , very thick] (\r+2.2,\s+0.5)--(\r+3,\s+.5);
		\node [above] at (\r+2.6,\s+0.5) {$X_{(f)}^n$};
		
		\draw [very thick] (\r+3,\s-0.5) rectangle (\r+5,\s+1.5);
		\node [align=center] at (\r+4,\s+0.5) {$W_{(f)}(\mathcal{Y}|\mathcal{X})$};
		
		\node [above] at (\r+5.5,\s+0.5) {$Y_{(f)}^n$};
		\draw [-> , very thick] (\r+5,\s+0.5)--(\r+6,\s+.5);
		
		\draw [very thick] (\r+6,\s-0.5) rectangle (\r+8,\s+1.5);
		\node [align=center] at (\r+7,\s+0.5) {Feedback Assisted\\Decoder};
		
		\draw [-> , very thick] (\r+8,\s+0.5)--(\r+9,\s+.5);
		\node [above] at (\r+9,\s+0.5) {$\hat{m} \in \mathcal{M}$};
		
		\draw [->, very thick] (\r+7,\s-0.5)--(\r+7,\s-2.5)--(\r+5,\s-2.5);
		
		\draw [very thick] (\r+3,\s-3.2) rectangle (\r+5,\s-1.8);
		\node [align=center] at (\r+4,\s-2.5) {$W_{(b)}(\mathcal{Y}|\mathcal{X})$};			
		
		\draw [->, very thick] (\r+3,\s-2.5)--(\r+1.3, \s-2.5)--(\r+1.3, \s-0.5);
		
		\node [above] at (\r+2.6,\s-2.5) {$Y_{(b)}^n$};			
		\node [above] at (\r+5.5,\s-2.5) {$X_{(b)}^n$};		
		
		\end{tikzpicture}
	}
	\caption{Communication Scheme for a DMC with active noisy feedback}
	\label{Fig:Noisy_Feedback}
\end{figure}
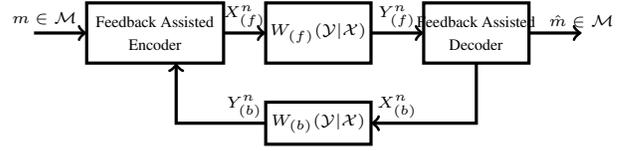

{\bf Adaptive zero-error capacity with feedback, $C_{0fa}$.} 
%{\color{red}These definitions need to be cleaned up. You define a zero-error adaptive code as a set of messages, sequence of encoding and decoding functions such that zero error is achieved. I do NOT think at channel use $j$ you need to produce a message output. You just need to produce your feedback signal. I think your definition here already assumed too much about how transmission takes places -- i.e. we do not NEED to repeat until we get a clean message, that is just one achievability scheme. Your definitions should not assume anything about how encoding or decoding works, just what inputs / outputs these functions can be a function of.   Please clean up the definitions. }
An adaptive zero-error code with feedback ${\cal C}_{0fa}({\cal M})$
 %{\color{red}N})$ \textcolor{red}{according to your definition random variable N should be used.} 
 for a DMC with a forward channel $(\mathcal{X}_{(f)}, \mathcal{Y}_{(f)}, W_{(f)})$ and a backward channel $(\mathcal{X}_{(b)}, \mathcal{Y}_{(b)}, W_{(b)})$
as in Fig. \ref{Fig:Noisy_Feedback} consists of a message set ${\cal M}$, a set of encoding ${\cal F}$, feedback ${\cal G}$, and decoding functions ${\cal H}$ respectively for $j=1,2,\cdots $ 
\begin{align*}
\mathcal{F}_{j}:& \mathcal{M} \times \mathcal{Y}_{(b)}^{j-1} \rightarrow \mathcal{X}_{(f)}, \\
\mathcal{G}_{j}:& \mathcal{Y}_{(f)}^{j-1}  \rightarrow \mathcal{X}_{(b)} \cup \emptyset \\
\mathcal{H}_{j}:& \mathcal{Y}_{(f)}^{j} \rightarrow \mathcal{M} \cup {\cal E}
\end{align*}
where, for every $m\in {\cal M}$, after some $N_m$ channel uses,  $H_{N_m}(Y_{(f)}^{N_m}) = m$  (i.e. zero-error after $N_m$ channel uses).
%\textbf{(I think $N_m$ should be for a specific message $m$ not all of them, and $N=\max\limits_{m\in \mathcal{M}} N_m$ is for the channel use for all messages.)}. 
At each channel use $j$, the transmitter sends ${\cal F}_j(m, y_{(b)}^{j-1})$ over the 
forward channel $W_{(f)}$, which is received and added to the sequence of received signals $y_{(f)}^j$. The receiver takes this sequence and transmits ${\cal G}_j(y_{(f)}^{j-1})$  back over the backwards channel, where it is received 
as $y_{(b),j}$. 
%If after $k$ channel uses the message $m$ is received with zero error for the first time, then we call $L_n(m) = k$ the delay for sending message $m$. 
%Let $N_m$ be the minimum delay for sending message $m$ with zero-error. 
Let $D_m\geq N_m$ be the number of channel uses needed for the message $m$ to be transmitted and decoded with zero error before the next message starts, and is a random variable.
Define the expected delay {per information bit}, with expectation taken over messages $m$ and channel instances,   as $\bar{D} := E[D_m]/\log|{\cal M}|$. The zero-error adaptive feedback capacity of the channel in Fig. \ref{Fig:Noisy_Feedback} is then given by the largest expected rate defined as $\bar{R} : = \frac{\log_{2} |{\cal M}|}{E[D_m]}$ such that there exists an adaptive zero-error code with feedback with expected delay $\bar{D} <\infty$.
% \textbf{(The problem with this definition is that $\bar{D}=1/\bar{R}$. The finite average delay $\bar{D}$ means that the average rate $\bar{R}$ should not be zero. I still think that this is not very general. Massey looks at the number of feedback channel use per one bit information bit on the forward channel as an extra cost for transmission over forward channel. The massey definition completely make sense especially for two-way channel. In fact, for two scheme with same average rate, the one that uses feedback channel less is the better scheme. However, in this definition, we assume that the feedback channel is always used with might not be true in general.  )}. 

 We use the  notation $C_{0fa}$ and $C_{0fa}^{noisy}$ to distinguish the zero-error adaptive feedback capacities when the feedback is noiseless and noisy, respectively. 

\section{Adaptive zero-error communication for channels with feedback}
\label{Sec:3-adaptive zero-error}
One way of ensuring zero-error communication in a channel with perfect feedback is to keep repeating a message until it is correctly received (the transmitter can verify correct reception from the perfect output feedback). 
One needs to then calculate the average rate and delay incurred. In the following, we present such communication schemes for channels with perfect (Theorem 1) and noisy (Theorem 2) feedback. We denote $x_i^j \bumpeq x'$ if there exists at least one $k \in [i, j]$ such that $x_k = x' $ (e.g. $1101 \bumpeq 0$). Let $[x_i]^l$ denote a sequence of $l$ repetitions of letter $x_i$ in some alphabet ${\cal X}$, $[x_i]^l = (x_i, x_i, \dots, x_i), |[x_i]^l|=l$.

\subsection{Complete, noiseless feedback}
 \label{Subsec:3-complete feedback}
 When complete, noiseless feedback is available, Massey \cite{Massey07} suggested a method for achieving zero-error at an expected rate approaching the small error or Shannon capacity of the forward channel, $C$. We outline a Theorem that we attribute to Massey below.  Let $\gamma_n = o(n)$, and $\gamma_n \rightarrow \infty$ as $n \rightarrow \infty$ (e.g. $\gamma_n=\log (n)$). Since the backward channel is noiseless, we omit subscript $_{(f)}$ for the forward channel and use $(\mathcal{X}, \mathcal{Y}, W)$. 
 
%{\color{red}Express Massey and Lapidoth Results and briefly explain how the data is transmitted.}
%%When complete feedback is available (Fig. \ref{Fig:Complete_Feedback}), the adaptive zero-error transmission is done as following:
%%
%%In each The decoder function is similar to $\mathcal{G}$

	\begin{algorithm}
		\label{Alg:ZAEF_complete}
		\SetKwInOut{Input}{Input}
		\SetKwInOut{Output}{Output}
		
 		\underline{Feedback Assisted Encoder}\;
 		\Input{$M \subseteq \mathcal{M},\mathcal{C}(\mathcal{M},n), \gamma_n$,   disprover triplet $(x_c,x_e, y_c)\in W $}
 		\Output{$L_{n}(m)$} %\tcc{$L_n$ is $\#$ of iterations}
		
		\ForAll{$m \in M$ }
		{
			$x^{n}\leftarrow c^{(n)}(m)$, $I \leftarrow 0$, $L_{n}(m) \leftarrow 0$    \;
		
			\While{$I=0$}
			{
				$L_{n}(m) \leftarrow L_{n}(m) + 1$ \;
				%\tcc*{$L_n$-th transmition iteration}
				Send $x^{n}$ through channel \;
				$\hat{m}={\cal G}(y^{n}(c^{(n)}(m)))$ \;
				\tcc*{$L_n$-th verification iteration} 
				\eIf{$\hat{m}\neq m$}
					{
						$x^{\gamma_n} \leftarrow [x_{e}]^{\gamma_n}$ \;
					}
					{
						$x^{\gamma_n} \leftarrow [x_{c}]^{\gamma_n}$ \;
					}
				Send $x^{\gamma_n}$ through channel \;
			
				\If{$y(x^{\gamma_n}) \bumpeq y_c$ }
				{
					$I \leftarrow 1$ \;
				}
			
		    }
    	}
    
		\caption{Adaptive zero-error communication scheme with complete feedback \cite{Massey07}}
\end{algorithm}

%\begin{figure}
%	\centering
%	{\scalefont{0.6}
%		\begin{tikzpicture}[xscale=0.7,yscale=0.4 ]
%		\def \r {6};
%		\def \s {14};
%		
%		
%		
%		\draw [-> , very thick] (\r-1.6,\s+0.5)--(\r-0.6,\s+.5);
%		\node [align=center] at (\r-1.4,\s+1) {$m \in \mathcal{M}$};
%		
%		\draw [very thick] (\r-0.6,\s-0.5) rectangle (\r+2.2,\s+1.5);
%		\node [align=center] at (\r+.8,\s+0.5) {Feedback Assisted\\Transmitter};
%		
%		\draw [-> , very thick] (\r+2.2,\s+0.5)--(\r+3,\s+.5);
%		\node [above] at (\r+2.6,\s+0.5) {$X^n$};
%		
%		\draw [very thick] (\r+3,\s-0.5) rectangle (\r+5,\s+1.5);
%		\node [align=center] at (\r+4,\s+0.5) {$W(\mathcal{Y}|\mathcal{X})$};
%		
%		\node [above] at (\r+5.5,\s+0.5) {$Y^n$};
%		\draw [-> , very thick] (\r+5,\s+0.5)--(\r+6,\s+.5);
%		
%		\draw [very thick] (\r+6,\s-0.5) rectangle (\r+7.5,\s+1.5);
%		\node [align=center] at (\r+6.75,\s+0.5) {Receiver};
%		
%		\draw [-> , very thick] (\r+7.5,\s+0.5)--(\r+8.5,\s+.5);
%		\node [above] at (\r+8.5,\s+0.5) {$\hat{m} \in \mathcal{M}$};
%		
%		\draw [->, very thick] (\r+6.75,\s-0.5)--(\r+6.75,\s-1.5)--(\r+1.3,\s-1.5)--(\r+1.3, \s-0.5);			
%		
%		\end{tikzpicture}
%	}
%	\caption{Communication Scheme for a DMC with noiseless feedback}
%	\label{Fig:Complete_Feedback}
%\end{figure}

\begin{Theorem}[Massey (Elaborated) \cite{Massey07}]
		The adaptive zero-error capacity $C_{0fa}$ for a DMC channel $({\cal X}, {\cal Y}, W)$ with noiseless feedback is
		\begin{equation}
		C_{0fa} = 
		\begin{cases} 
		C &\mbox{if } C_{0u}>0  \\ 
		0 & \mbox{otherwise,} 
		\end{cases}
		\end{equation}
		where $C$ denotes the Shannon capacity of  the channel $({\cal X}, {\cal Y}, W)$, and $C_{0u}$ denotes its zero-undetected error capacity.
		\end{Theorem}
\begin{Proof}
		 If $C_{0u}=0$ then by \cite{Forney68}, channel $W$ does not have a disprover, i.e. for every $x \in \mathcal{X}, y \in \mathcal{Y}, W(y|x)>0$. Thus, no matter which sequence is sent 
		 %during verification step ($x_1^{\gamma_n}$ in Alg. \ref{Alg:ZAEF_complete}), 
		 the receiver is unable to decide anything with zero error and $C_{0fa}=0$. %. The presence of feedback cannot help. 
		 %; Thereby, the receiver never verifies the correct transmission of first $x_1^n$. Hence, $C_{0fi}=0$.
		 
		 When $C_{0u}>0$, we may assume that the DMC $W$ contains at least one disprover. Equivalently, there exists at least one triple $(x_c,x_e,y_c) \in (\mathcal{X} \times \mathcal{X} \times \mathcal{Y})$ such that $W(y_c|x_e)=0$ and $W(y_c|x_c)>0$. 
		 
		 %	 The converse proof is trivial using the fact that feedback does not increase the Shannon capacity and $C_{0fi}\leq C_{fi} \leq C$. 
		 %	 \nrd{What is $C_{fi}$? Not defined -- I think you mean the adaptive capacity of a channel with perfect feedback? In this case, I would call is $C_f$, and the statement would be: ``
		 The converse proof is trivial using $C_{0fa}\overset{(1)}{\leq} C_f \overset{(2)}{=} C$, where (1) follows as $C_f$ denotes the small-error capacity of the channel with perfect feedback, which is always an outer bound to the more restrictive zero-error setting, and (2) follows from Shannon's result that perfect feedback does not increase the small error capacity of a channel.
		 
		 For the achievability, let $\mathcal{C}(\mathcal{M},n)$ be a capacity achieving code for the DMC $W$ whose maximal probability of error $\lambda^{(n)}$ tends to zero and whose rate approaches the Shannon capacity $C$ as block length $n \rightarrow \infty$. Note that the output block $y^n$ is available in real time at the transmitter due to the presence of perfect feedback. The transmitter can thus mimic the receiver's decoding rule and determine whether the receiver obtained the correct message. It then tells the receiver this by sending $\gamma_n$ copies of either $x_c$ (if correct) or $x_e$ (if erroneous) through the noisy $W$. Since the receiver can only receive a $y_c$ from an $x_c$ (definition of a disprover), once it receives at least one $y_c$ it realizes that its decoded message is correct, and zero-error communication is achieved. We note that variable $I$ in the Algorithm \ref{Alg:ZAEF_complete} is used to synchronize the transmitted and receivers, i.e. indicates when a new message will start.
		 % It is easy to verify that this ``toggle bit'' as Massey called it \cite{Massey07} indeed allows transmitter and receiver to agree upon when a codeword is new versus repeated.}{\color{green} The idea of toggle bit in massey was used when the feedback is noisy. I am not sure if this is same as toggle bit, but I agree that they are both for synchronization.} %when it is a repetition of the previous not correctly decoded one.}
		 % It resends the same codeword until at 
		 % \emph{feedback assisted transmitter module} (Fig. \ref{Fig:Complete_Feedback}) due to the complete feedback. Thus, this module can do the same decoding process and figure out the same message $\hat{m}$ that receiver decodes (line 5 in algorithm \ref{Alg:ZAEF_complete}). The transmitter will resend the codeword $c_n(m)$ until 
		 %		 at least one $y_c$ is received during verification stage ($y(x_{1}^{\gamma_n}) \bumpeq y_c$). Since $y_c$ is the disprover for $W_d$, it is impossible to receive $y_c$ by transmitting $x_e$. Thus, the output letter $y_c$ is received if, and only if, $\hat{m}=m$, and thus the zero-error transmission $(\hat{m} = m)$ is guaranteed
		 %		 \[
		 %		 P_e = Pr(\hat{m} \neq m | m \mbox{ was sent}) = 0.
		 %		 \]

		 To calculate the average rate and delay achieved, note that the probability that a message is correctly received with zero error is the probability that the message was correctly received at the receiver after seeing the codeword of length $n$, and then the receiver seeing at least one correct indicator (i.e. seeing one $y_c$) in a block of length $\gamma_n$. Hence after $n+\gamma_n$ channel uses, the probability of correctly decoding message $m$ is $p_{n,m} =  (1-\lambda_m^{(n)}) \bigg[1-\big(1-W(y_c|x_c)\big)^{\gamma_n} \bigg]$. Viewing this as a probability of success, the number of codeword re-transmissions needed to correctly receive message $m$ and wait for the transceivers to synchronize and start a new message is hence a geometric random variable $L_n(m)$ with $E[L_n(m)]=1/p_{n,m}$.

%		 let $L_n(m)$ denote the number of retransmissions of length $n$ needed to successfully transmit message $m$. Let $p_{n,m}^{(j)}$ 
%		 be the probability of successfully transmitting the message $m \in \mathcal{M}$ using capacity achieving code $\mathcal{C}(\mathcal{M},n)$ in j-th iteration ($L_n(m)=j$). 
%		 \nrd{Fix notation in the 
%		 	below and in algorithms -- why are there subscripts $1$? I think it is confusing, just remove the 1 if you agree.  Also, is $W^{\gamma_n}(y^{\gamma_n} \bumpeq y_c | X_1^{\gamma_n}=[x_c]^{\gamma_n})$ defined?}
%		 \begin{eqnarray}
%		 p_{n,m}^{(j)} 
%		 &\triangleq& Pr(I = 1 | L_n = j ) \nonumber\\ 
%		 &=& Pr(\hat{m} = m | L_n = j ) \nonumber\\
%		 && \quad \times W^{\gamma_n}(y^{\gamma_n} \bumpeq y_c | X_1^{\gamma_n}=[x_c]^{\gamma_n})\nonumber\\
%		 &=& (1-\lambda_m^{(n)}) \bigg[1-\big(1-W(y_c|x_c)\big)^{\gamma_n} \bigg]. 		 
%%		 &=& (1-P_{e}^{(n)}) \bigg[1-\big(1-W(y_c|x_c)^{\gamma_n} \big)\bigg]. 
%		 \end{eqnarray}
%		 Note that $p_{n,m}^{(j)}$ does not depend on value of $j$. Thus, for the sake of simplicity, we denote $p_{n,m} \triangleq p_{n,m}^{(j)}$. Since $L_n(m)$ is increased until the first successful transmission occurs, the random variable $L_n(m)$ tends to have geometric distribution with probability of success ($p_{n,m}$). Hence, $E[L_n(m)]=1/p_{n,m}$. Note that for this scheme
		 
		 Hence, the delay incurred to correctly decode message $m$ is 
		 $N_m = (n+\gamma_n)\cdot L_n(m)$ and hence the expected delay (not yet normalized by the number of bits) is 
%		 Let $M \in \mathcal{M}$ denote the random message that is sent. Thus,
		 \begin{eqnarray}
		 \bar{N} 
		 & \stackrel{(1)}{=}& E[E[N_m]]   = E[E[(n+\gamma_n) \cdot L_n(m)]]\nonumber\\
		 %&=& E[E[N|M=m]] = E[E[N_m]]\nonumber\\
		 &=& (n+\gamma_n)\cdot E\left[\frac{1}{p_{n,m}}\right] \stackrel{(2)}{=} \frac{n+\gamma_n}{|\mathcal{M}|} \sum\limits_{m=1}^{|\mathcal{M}|} \frac{1}{p_{n,m}}, 
		 \end{eqnarray}
		 where in (1), the outer expectation is with respect to the uniform distribution over messages $m\in {\cal M}$ and the inner expectation is over the channels, and where $(2)$ is because message $m$ is uniform over ${\cal M}$. Hence, as $n\rightarrow \infty$
		 \begin{eqnarray}
		 \lim_{n\rightarrow \infty} \bar{R} 
		 &=& \lim_{n\rightarrow \infty} \frac{\log_{2} |\mathcal{M}|}{\bar{N}}\nonumber\\
		 &=& \lim_{n\rightarrow \infty} \frac{\log_{2} |\mathcal{M}|}{n} \frac{|\mathcal{M}|}{(1+\frac{\gamma_n}{n}) \sum\limits_{m=1}^{|\mathcal{M}|} (p_{n,m})^{-1} } \overset{(2)}{=} C \nonumber
		 \end{eqnarray}
		   where $(2)$ follows as we are using a Shannon capacity achieving code  $\mathcal{C}(\mathcal{M}, n)$,  and by definitions of $\gamma_n$ and $p_{n,m}$. 
%		 {\color{red} Note that the average rate $\bar{R}$ is a function of $n$.} Since capacity acheiving code $\mathcal{C}(\mathcal{M}, n)$ has the rate equals $C$ as $n \rightarrow \infty$, the average rate of $\bar{R}$ also converges to $C$ as $n \rightarrow \infty$. 
The expected delay incurred is $\bar{D} := \lim_{n\rightarrow \infty} E[N_m]/\log|{\cal M}| = \lim_{n\rightarrow\infty} \frac{n+\gamma_n}{nC} = \frac{1}{C} < \infty$ as needed.
%		 		 In this scheme one bit feedback is needed per each bit transmission on forward channel $W$. Thus, {\color{red} do we need $\bar{N}$ to be finite or $\bar{D}$ to be finite? I am still confused about these definitions!}
%		 \[
%		 D_{n}(m) = \frac{n+ \gamma_n}{\log_{2} |\mathcal{M}|},
%		 \]
%		 Thus, the average feedback delay $\bar{D}= 1/C$ as $n \rightarrow \infty$. 
%		     
\end{Proof}

%\begin{Remark}
%
%\end{Remark}

%\begin{Remark}
%	Explain the mismatch issue and why we need to transmit $\gamma_n$ times letter $x_e$. Also explain the data interleaving to not let the transmitter to stay idle.
%	
%\end{Remark}
%
%\begin{Remark}
%	Address Amos paper and tell that their proof technique is similar to Massey's paper. However, the proof is for $C_{0fu}$.
%\end{Remark}

%An adaptive-zero-error for 
%
%We use a flagged-zero-undetected-error (FZUE) code of block length n  to transmit the data through direct channel. After receiving $y_{1}^{\gamma_n} \in \mathcal{Y}_f^{\gamma_n}$, the FUZE encoding function
%\begin{equation}
%f: \mathcal{B} \times \mathcal{M} \rightarrow \mathcal{X}_{d}^{n},
%\end{equation}
%encode the message. Note that $\mathcal{B}=\{0,1\}$ and $\mathcal{M}$ denote the message set. After receiving $y_{1}^n \in \mathcal{Y}^{n}$, the decoder lists all probable messages that might generate the corresponding output $y_1^n$. Let $M(y_{1}^{n})$ denote the set of probable messages corresponding to $y_{1}^{n}$

\subsection{Noisy feedback}
\label{Subsec:4-noisy_feedback}
%In previous section, we discussed the problem of adaptive-zero-error communication over a DMC $W_{d}$ in the presence of noiseless feedback. Because the feedback is  noiseless, after each block transmission the transmitter is able to verify whether the decoded message is error free or not. In fact, the decoding process is concurrently done at both receiver and transmitter. If transmitter can not verify that the decoded message is correct, the transmitter resend the same code again. This process is repeated until not only transmitter be sure that the perfect decoding was done at receiver, but the transmitter also successfully notify the receiver that the received data is correct during verification process. Therefore, this scheme eventually provides communication with zerro-error. 

When the feedback channel is noisy,  the above scheme no longer works as i) the transmitter does not have perfect access to the received signal, and hence cannot mimic the decoding process. It is thus harder to ensure zero error; and  ii) synchronizing the transmitter and receiver becomes more challenging as both channels are noisy. How can the receiver know when a codeword is new versus when it is repeated?
% the transmitter and receiver might become unsynchronized when transmitter repeat a message while the receiver expect a new message. 
When feedback is noiseless, the synchronization issue can be completely resolved at the transmitter.  %However, in the noisy feedback case this problem cannot be resolved only at the transmitter, and a
With noisy feedback, we propose a new synchronization technique. % using states $s_t, s_r$. %a novel synchronized feedback assisted encoder and receiver are needed.

In \cite{Massey07} an adaptive zero-error communication scheme for DMC with noisy feedback was proposed. The synchronized feedback assisted transmitter and receiver are described using Algorithms \ref{Alg:ZAEF_Trans_Noisy} and \ref{Alg:ZAEF_Rec_Noisy}, respectively.  {In these,  $s_t, s_r \in \mathcal{B}$ are the current states of the transmitter and receiver respectively. When equal, both transmitter and receiver are working on transmitting a new message; when different, the receiver has decoded the message but the transmitter does not know this yet due to the noisy feedback channel.}
	\begin{algorithm}
	\label{Alg:ZAEF_Trans_Noisy}
	\SetKwInOut{Input}{Input}
	\SetKwInOut{Output}{Output}
	
	\underline{Synchronized Feedback Assisted Transmitter} (Fig. \ref{Fig:Noisy_Feedback})\;
	\Input{$m \in \mathcal{M},\mathcal{C}_{0u}^{(f)}(\mathcal{M},n), \gamma_n, y_c \in \mathcal{Y}_{(b)} $}
	\Output{$L_{n}(m)$} %\tcc{$L_n$ is $\#$ of iterations}
	
	$s_t \leftarrow 0$ 	\tcc{Transmitter state}
	 
	\ForAll{ $m$ that need to be sent }
	{
		$b_1 \leftarrow s_t$\;
	$(b_2,b_3, \cdots, b_{k}) \leftarrow m$ \;
	$x^{n}\leftarrow c_{0u}^{(n)}(b_{1}^{k})$, $I \leftarrow 0$, $L_{n}(m) \leftarrow 0$    \;
	
	\While{$I=0$}
	{
		$L_{n}(m) \leftarrow L_{n}(m) + 1$ \;
		
		\tcc*{$L_n$-th transmition Stage }
		Send $x^{n}$ through channel \;
		$\hat{m}=\mathcal{G}_{0u}(y_{1}^{n}(c_{0u}^{n}(m)))$ \;
		\tcc*{$L_n(m)$-th verification stage} 
%		\eIf{$\hat{m}\neq m$}
%		{
%			$x_{1}^{\gamma_n} \leftarrow (x_{e}, x_{e}, \dots, x_{e})$ \;
%		}
%		{
%			$x_{1}^{\gamma_n} \leftarrow (x_{c}, x_{c}, \dots, x_{c})$ \;
%		}
		Receive $y_{1}^{\gamma_n}$ through feedback channel \;
		
		\If{$y_{1}^{\gamma_n} \bumpeq y_c$}
		{
			$I \leftarrow 1$ \;
		}
		
	}
	$s_t \leftarrow s_t^{c}$ \tcc{Inform receiver about new message}
	}
	\caption{Adaptive zero-error communication scheme with Noisy feedback}
\end{algorithm}

	\begin{algorithm}
	\label{Alg:ZAEF_Rec_Noisy}
	\SetKwInOut{Input}{Input}
	\SetKwInOut{Output}{Output}
	
	\underline{Synchronized Feedback Assisted Receiver} (Fig. \ref{Fig:Noisy_Feedback})\;
	\Input{$y^{n} \in \mathcal{Y}_{(f)}^{n}, \gamma_n, (x_e, x_c) \in \mathcal{X}_{(b)} $ for $W_{(b)}$}
	\Output{$(\hat{b}_{2}^{k})$ } %\tcc{$L_n$ is $\#$ of iterations}
	
	$s_r \leftarrow 0$ 	\tcc{Receiver state} 
%	$\hat{b}_2^k=[0]^{k-1}$\;
%	\tcc{Received message}
%	
	\ForAll{ Received $y^{n}$}
	{
			
	%$M(y^{n})= \{m \in \mathcal{M}: W_{(f)}(y^n|\mathcal{F}_{0}^{(0u)}(m))>0 \mbox{ or } W_{(f)}(y^n|\mathcal{F}_{1}^{(0u)}(m)) >0\}$ \;
	$M(y^{n})= \{m \in \mathcal{M}: W_{(f)}(y^n|c^{(n)}_{0u}(m))>0 \mbox{ or } W_{(f)}(y^n|c_{0u}^{(n)}(m)) >0\}$ \;
	
	\eIf{$|M(y^{n})|=1$}
	{
		$\hat{b}_{1}^{k} \leftarrow M(y^{n})$ \; 
		\If{$\hat{b}_{1} = s_r$}
		{			
			$\hat{m} \leftarrow \hat{b}_{2}^{k}$ 		\tcc{Store message $\hat{m}$} 
			$s_r \leftarrow s_r^{c}$ \;
			\tcc{Complement $s_r$ to indicate ready for new message} 
		}
		$x^{\gamma_n} \leftarrow [x_{c}]^\gamma_n$ \;		
	}
	{
		$x^{\gamma_n} \leftarrow [x_{e}]^\gamma_n$ \;		
	}
	Send $x^{\gamma_n}$ through feedback channel $W_{(f)}$ \;	
	}
	\caption{Iterative zero-error communication scheme with Noisy feedback}
\end{algorithm}

\begin{Theorem}
	The adaptive zero-error capacity of a forward DMC $W_{(f)}$ with noisy feedback DMC $W_{(b)}$ shown in Fig. \ref{Fig:Noisy_Feedback}, denoted by $C_{0fa}^{noisy}$, satisfies %{\color{red} ND DISAGREES -- CAN HAVE $C_{0u}^{(b)} = 0$ and still have a positive forward zero error capacity as long as $\alpha(G_{X|Y}^{(f)}) > 1$. So I removed the condition! equals $0$ if either $C_{0u}^{(f)}=0$ or $C_{0u}^{(b)}=0$.} %Furthermore, 
	\begin{equation}
	C_{0fa}^{noisy} \geq C_{0u}^{(f)} \mbox{\quad if } C_{0u}^{(f)}>0 \mbox{ and }  C_{0u}^{(b)}>0, \label{eq:noisy}
	\end{equation}
	where $C_{0u}^{(f)}$ and $C_{0u}^{(b)}$ denote the zero-undetected error capacities of the forward and backward links. 
	If furthermore, for some positive functions $A(\cdot)$ and $B(\cdot)$ and some capacity-achieving input distribution $Q^{*}$, $W_{(f)}(y|x)=A(x)B(y)$ holds whenever $Q^{*}(x)W_{(f)}(y|x)>0$, then
	$C_{0fa} = C^{(f)}$.
	\end{Theorem}
%	\begin{equation}
%	C_{0fa}^{noisy} = C^{(f)} \mbox{\quad if } C_{0u}^{(f)}>0 \mbox{ and }  C_{0u}^{(b)}>0.  \label{eq:converse}
%	\end{equation}	  
\begin{Proof}
	%***************** NO LONGER NEED THIS AS REMOVED THE CLAIM ABOUT WHEN POSITIVE ************************************
%	If $C_{0u}^{(f)} = 0$, Massey \cite{Massey07} showed that $C_{0fi}^{perfect}=0$ even for noiseless feedback. Hence, $C_{0u}^{(f)} > 0$ is necessary for achieving positive $C_{0fi}^{noisy}>0$. If $C_{0u}^{(b)} = 0$, meaning that there exists no disprover $y_c \in \mathcal{Y}_f$ for feedback channel $W_{(b)}$. Thus, no matter which input sequence is used for verification purpose, the transmitter is not able to rule out any of these sequences. Therefore, it is impossible to send any message with zero error even if the message is retransmitted for an infinite period.
Since $C_{0u}^{(b)}$ is positive,  there exists at least one triple 
%$(x_c,x_e,y_c) \in (\mathcal{X}_{(f)} \times \mathcal{X}_{(f)} \times \mathcal{Y}_{(f)})$ such that $W_{(f)}(y_c|x_e)=0$ and $W_{(f)}(y_c|x_c)>0$. Similarly, there exists at least one triple 
$(x'_c,x'_e,y'_c) \in (\mathcal{X}_{(b)} \times \mathcal{X}_{(b)} \times \mathcal{Y}_{(b)})$ such that $W_{(b)}(y'_c|x'_e)=0$ and $W_{(b)}(y'_c|x'_c)>0$. 
Note that we require $C_{0u}^{(f)}$ to be positive as well, else no zero-error communication can take place at all, not even with perfect feedback.
	
For achievability of \eqref{eq:noisy}, take a zero-undetected-error capacity achieving code $C_{0u}(\mathcal{M}, n)$ for channel $W_{(f)}$ whose maximal erasure probability tends to zero and whose rate approaches $C_{0u}^{(f)}$. Note that the first message bit  $b_1$ out of the bit stream of length $k, b_1^k$ (that is encoded) carries the transmitter's state variable $s_t$, used for synchronization. 
%. Here $k = n C_{0u}^{(f)}$. This will be used for synchronization purposes.
%Since $C_{0u}{(f)}>0$ such a code exists. 
%Suppose we would like to transmit $m\in \mathcal{M}$, and let $c_{0u}^{(n)}(m) \in C_{0u}(\mathcal{M}, n)$ be the length-$n$ codeword corresponding to message $m$. The 
%{\color{red} Need to state that the first bit of the message is the state!}

To transmit message $m\in {\cal M}$, codeword $c_{0u}^{(n)}(m)$ is sent through $W_{(f)}$. Upon receiving $y^n \in \mathcal{Y}_{(f)}^{n}$, the zero-undetected-error decoder is used to obtain an estimate of the message. 	Since the probability of undetected-error is equal to zero, 
%{\color{red} as $n\rightarrow \infty$?} ({\color{green} $n \rightarrow \infty$ is not needed for probability of undetected error, This is used to make probability of erasure converging to zero.}),
 the only type of error that might occur is an erasure ($|M(y^n)|>1$, see \eqref{eq:probable}). If there is an erasure, according to Algorithm 3, the receiver informs the transmitter by sending $\gamma_n$ repetitions of the letter $x'_e$ (i.e. it sends $[x'_e]^{\gamma_n}$). Since $W_{(b)}(y'_c|x'_e)=0$, it is impossible to  receive $y'_{c}$ at the transmitter through the noisy feedback channel. Thus, the transmitter -- not seeing any $y_c'$ -- again transmits $c_{0u}^{(n)}(m)$. 
%As long as the receiver declares erasure, the original codeword $c_{0u}^{(n)}(m)$ is re-transmitted. 

%	{\color{red} I guess this part will go into a new lemma? I think we could just say {\bf Ensuring the transmitter and receiver remain synchronized.}  We assume that transmitter and receiver are synced at the start of transmission (WLOG $s_t = s_r = 0$). 
%	
	
A message is re-transmitted until the following happens. In the first iteration that $|M(y^n)| = 1$, the receiver sets $\hat{m}=M(y^n) = b_2^k$ (recalling that the first bit $b_1$ carries the state $s_t$ and not the message), and knows with probability 1 that this is the correct message (i.e. zero-error in decoding the message by definition of a zero undetected error code). The challenge now is to tell the transmitter, through the noisy channel, that it has received the message and hence that the transmitter can move on to a new message.
This is done through a careful protocol for keeping the binary states $s_t$ and $s_r$ at the transmitter and receiver synchronized.   They start off synchronized to $b_1$. 
%We assume that transmitter and receiver are synced at the start of transmission ($s_t = s_r = b_1$). 
% Let $\hat{m}=\hat{b}_{1}^{k}$. 
%The transmitter stores $\hat{b}_{2}^{k}$ as the transmitted message. 
Once the receiver sees $|M(y^n)| = 1$, and looks at the decoded message, before sending confirmation that it received the message,  the receiver switches its internal state, i.e. $s_r = s_r^{c} = b_1^c$ (the complement of the first bit) . 
Then, it conveys correct decoding by repeating $x'_c$ $\gamma_n$ times through the feedback channel $W_{(b)}$. Two things can now happen: 

1) If the letter $y'_c$ is not received at the transmitter,  then the transmitter sends back the same message and the process repeats. At this stage then, $s_t = b_1$ while $s_r=b_1^c$. This process repeats until the decoder uniquely decodes $|M(y^n)|=1$ AND the state bits match. If the state bits do not match, the receiver does not update the decoded message. 
% However, since the transmitter and receiver are not synchronized in this stage ($s_{r}\neq s_t = \hat{b}_{1}$), the receiver does not store $\hat{b}_{2}^{k}$ as the new transmitted message. But, it still sends back $X_{1}^{\gamma_n}=[x'_c]^{\gamma_n}$ 
%The receiver continues to send back $\gamma_n$ copies of $x_c'$ to ask the transmitter for a new message. 
%The process is continuous until $y_{1}^{\gamma_n}([x'_c]^{\gamma_n}) \bumpeq y'_c$. In this case, transmitter changes its own state $s_t=s_t^c$, which makes receiver and transmitter are synced again before transmitting the new message.

2) If it does receive $y_c'$, then it knows the receiver successfully and uniquely decoded the message and hence it sets $s_t = s_t^c$. At this point then the transmitter and receiver states are again equal $s_t=s_r$. A new message, with the new state again as first bit, is transmitted.

%	Note that for every message $m\in \mathcal{M}$, the corresponding synced zero-undetected-error codeword  is iteratively transmitted until the output letter $y'_c$ is received at the transmitted through feedback channel $W_{(f)}$ which means that eventually $|M(y^n(c_{0u}^{(n)}(m)))| = 1$. Thus the zero-error transmission $(\hat{m} = m)$ is guaranteed
%	\[
%	P_e = Pr(\hat{m} \neq m | m \mbox{ was sent}) = 0.
%	\]

As in the previous case, to calculate the average rate and delay achieved, note that the probability that a message is correctly received with zero error is the probability that the message was correctly received at the receiver 
after seeing the codeword of length $n$, and then the {\it transmitter} (now through a noisy channel) seeing at least one $y_c'$ in a block of length $\gamma_n$. Hence after $n+\gamma_n$ 
channel uses, the probability of correctly decoding message $m$ is $p_{n,m} =  (1-\lambda_m^{(n)}) \bigg[1-\big(1-W_{(b)}(y_c'|x_c')\big)^{\gamma_n} \bigg]$. Viewing this as a probability of success, the number of codeword re-
transmissions needed to transmit message $m$ is hence a geometric random variable $L_n(m)$ with $E[L_n(m)]=1/p_{n,m}$. {The analysis of the achieved average rate and delay is identical to Theorem 1}, 
%\textbf{I think they are not completely the same. For $\bar{D}$ is different as}
% \[
%{\color{red} \bar{D} := \lim_{n\rightarrow \infty} \gamma_n E[L_{n}(m)]/\log|{\cal M}| = \lim_{n\rightarrow\infty} \frac{\gamma_n}{nC} = 0 < \infty}
% \]
%
except that we 
now use the backward $W_{(b)}(y_c'|x_c')$ in the definition of $p_{n,m}$, and the code we use is an undetected error capacity achieving code, in which case the rate tends to $C_{0u}^{(f)}$ as $n\rightarrow \infty$.

	To show that our bound is tight for the class of channels stated below \eqref{eq:noisy}, note that  Csisz\'ar and Narayan showed that if $C_{0u}^{(f)}>0$, and if the conditions after \eqref{eq:noisy} hold
	% for some positive functions $A(\cdot)$ and $B(\cdot)$ and some capacity-achieving input distribution $Q^{*}$, the relation $W_{(f)}(y|x)=A(x)B(y)$ holds whenever $Q^{*}(x)W_{(f)}(y|x)>0$, 
	then the zero-undetected capacity becomes equal to small error Shannon capacity ($C_{0u}^{(f)}=C^{(f)}$). Thus, for these channels,  we can easily prove that $C_{0fa}^{noisy} \geq C^{(f)}$. This is tight, as we always have $C_{0fa}^{noisy} \leq C_{0fa}\leq C^{(f)}$.  
\end{Proof}

\section{Conclusion}
A major difference between our adaptive-zero-error communication schemes with noiseless versus noisy feedback is that the verification sequence (i.e. transmitter and receiver agreeing the receiver has decoded it successfully) is sent by the transmitter in the noiseless case whereas it is sent by the receiver in the noisy case. In the former, the perfect feedback allows us to approach rates up to $C$ as undetected errors can be caught by the transmitter. In the latter, due to the noisy feedback, our scheme must backoff from $C$ to $C_{0u}$ in order to ensure that no undetected errors occur, as they cannot be corrected by the transmitter under our scheme. 

\section*{Acknowledgment}
The work of the authors was partially supported by NSF under award 1645381. The contents of this article are solely the responsibility of the authors and do not necessarily represent the official views of the NSF. 
%{\color{red} Fix Forney reference and send me updated bib file!}

% trigger a \newpage just before the given reference
% number - used to balance the columns on the last page
% adjust value as needed - may need to be readjusted if
% the document is modified later
%\IEEEtriggeratref{8}
% The "triggered" command can be changed if desired:
%\IEEEtriggercmd{\enlargethispage{-5in}}

% references section

% can use a bibliography generated by BibTeX as a .bbl file
% BibTeX documentation can be easily obtained at:
% http://www.ctan.org/tex-archive/biblio/bibtex/contrib/doc/
% The IEEEtran BibTeX style support page is at:
% http://www.michaelshell.org/tex/ieeetran/bibtex/
\bibliographystyle{IEEEtran}
% argument is your BibTeX string definitions and bibliography database(s)
\bibliography{mybibfile}

%
% <OR> manually copy in the resultant .bbl file
% set second argument of \begin to the number of references
% (used to reserve space for the reference number labels box)
%\begin{thebibliography}{1}
%
%%\bibitem{IEEEhowto:kopka}
%%H.~Kopka and P.~W. Daly, \emph{A Guide to \LaTeX}, 3rd~ed.\hskip 1em plus
%%  0.5em minus 0.4em\relax Harlow, England: Addison-Wesley, 1999.
%
%\bibitem{IEEEhowto:kopka}
%

%\end{thebibliography}

% that's all folks
\end{document}